\documentclass[sigconf]{acmart}
\AtBeginDocument{%
  }

\setcopyright{acmlicensed}
\copyrightyear{2026}
\acmYear{2026}
\acmDOI{XXXXXXX.XXXXXXX}
\acmConference[UMAP Adjunct '26]{Adjunct Proceedings of the 34rd ACM Conference on User Modeling, Adaptation
and Personalization}{June 03--05,
  2026}{UMAP, Sweden}
\acmISBN{978-1-4503-XXXX-X/2018/06}

\begin{document}

\title{Beyond Centralization: User-Controlled Federated Recommendations in Practice}

\author{Manel Slokom}
\email{manel.slokom@cwi.nl}
\orcid{0000-0002-9048-1906}
\affiliation{%
  \institution{Centrum Wiskunde \& Informatica}
  \country{The Netherlands}
}
\author{Alejandro Bellogin}
\orcid{0000-0001-6368-2510}
 \email{alejandro.bellogin@uam.es} 
\affiliation{%
  \institution{Universidad Autónoma de Madrid}
  \country{Spain}
}

\begin{abstract}
Recommendation systems typically require centralized user data, limiting user control and raising privacy concerns. 
Federated learning offers an alternative by keeping data on-device, but its impact on real user behavior remains largely unexplored.
We present a live federated recommender system that allows users to control the recommendation objective while keeping their data local. 
In a 53-day deployment with 22 participants and a catalog of 8807 titles, users interacted with recommendations and switched between personalization and diversity-enhanced ranking.
We find that users prefer personalization when given explicit choice (65.37\% vs.\ 62.07\% CTR), actively engage with control mechanisms (3.93/5 satisfaction; 248 settings changes), and develop an understanding of how their interactions affect recommendations through immediate feedback.
Our results show that user control, privacy, and effective personalization can be combined in a working system. 
We demonstrate a practical approach to interactive, privacy-preserving recommendation.
Code and demo materials are available at: \url{https://github.com/SlokomManel/federated-recommendations-participants}

\end{abstract}



\keywords{Federated learning, recommender systems, privacy-preserving personalization, user agency, user studies}


\maketitle

\section{Introduction}

Recommendation systems shape what people watch, read, and listen to. 
In most real-world deployments, however, users have little visibility into or control over what data is collected, how it is processed, and what the system optimizes for. 
This creates a gap between what systems do \emph{to} users and what users can do \emph{with} them.
This paper asks a simple question: can we give users direct control over recommendations while preserving both personalization and privacy?

To answer this, we build and deploy a \textbf{federated recommender system}. 
In federated learning, user data remains on-device; models are trained locally, and only model updates (not raw data) are shared and aggregated~\cite{fedrecsurvey}. 
This allows systems to learn from many users without centralizing personal data.

We implement this approach in a working system where participants receive movie recommendations, control the recommendation objective, customize the interface, and observe how their interactions affect results. 
Users can switch between (i) standard personalization and (ii) diversity-enhanced ranking, or opt out of aggregation entirely. 
In this design, privacy is enforced by architecture, and control is exposed through the interface.

We deploy the system with 22 participants over 53 consecutive days using a catalog of 8807 titles. 
While prior work has studied federated recommendation in offline or simulated settings~\cite{fedrecsurvey,FedRec,FDRS,FCF}, we focus on real user behavior in a live system, evaluating both recommendation outcomes and how users engage with control.

\textbf{Our contribution:} To the best of our knowledge, this is among the first live deployments of a federated recommender evaluated in a longitudinal study with participants interacting with the system over 53 days. 
Our system allows users to directly control the recommendation objective by switching between standard personalization and diversity-enhanced ranking. 
We focus on the following question: \textit{how do users exercise this control, and what preferences emerge between the two modes?} 
We show that (i) users prefer personalization when given explicit choice, (ii) they actively engage with control mechanisms, and (iii) they understand how their interactions affect recommendations when feedback is immediate. 
\section{System Overview and Interaction Design}

\begin{figure*}
    \centering
    \includegraphics[width=0.8\linewidth]{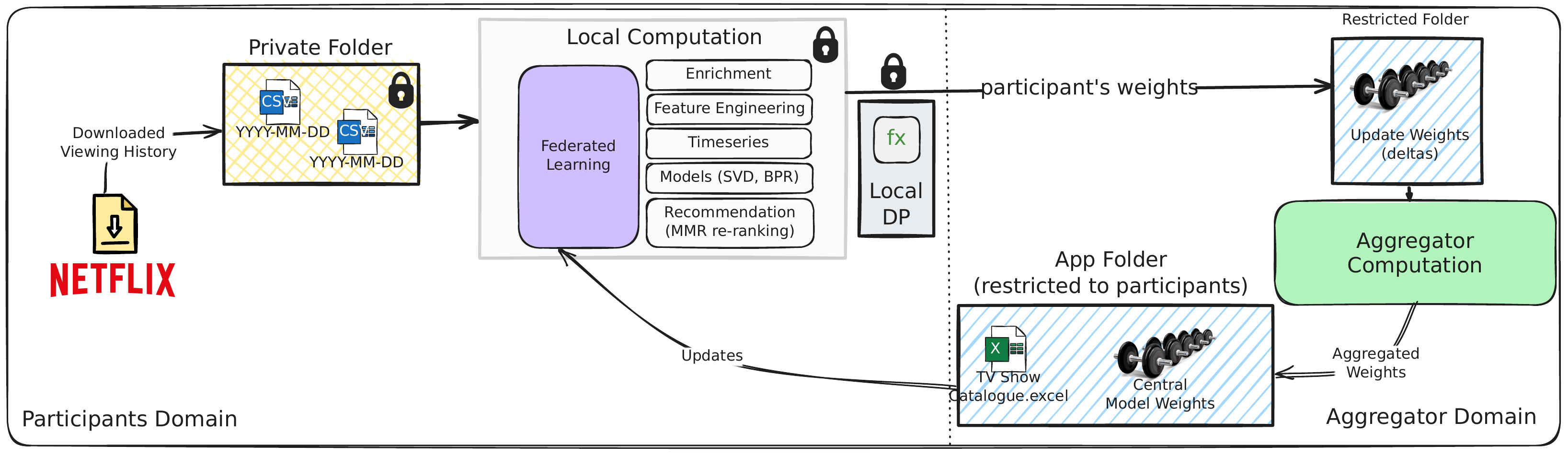}
    \caption{End-to-end demo loop. Viewing history stays on device. Local training produces model updates. We apply differential privacy ($\varepsilon = 2.0$) and share only updates. The aggregator combines updates and redistributes global parameters.}
    \label{fig:FedFlex}
\end{figure*}

Our system consists of two tightly coupled layers: (i) a recommendation pipeline and (ii) an interaction layer that exposes control to users. 
Figure~\ref{fig:FedFlex} illustrates the end-to-end loop.

\subsection{Recommendation pipeline}

\textbf{Local model training.} Each device trains a BPR (Bayesian Personalized Ranking) model on its viewing history. 
BPR optimizes ranking accuracy through pairwise loss and requires no centralized logging of user data~\cite{BPR}.

\textbf{Aggregation.} After local training, encrypted updates are sent to a central aggregator, which combines them by averaging model parameters across participants and redistributes the global model. 
At no point does the server access raw user data; only model updates are exchanged.

\textbf{Differential privacy.} We apply Gaussian noise ($\varepsilon = 2.0$) before aggregation to prevent preference inference from aggregated updates~\cite{Dwork2008Differential,gaussiandp}.

\subsection{Interaction design}
\label{sec:interaction}

\begin{figure*}[h]
    \centering
    \includegraphics[width=0.9\linewidth]{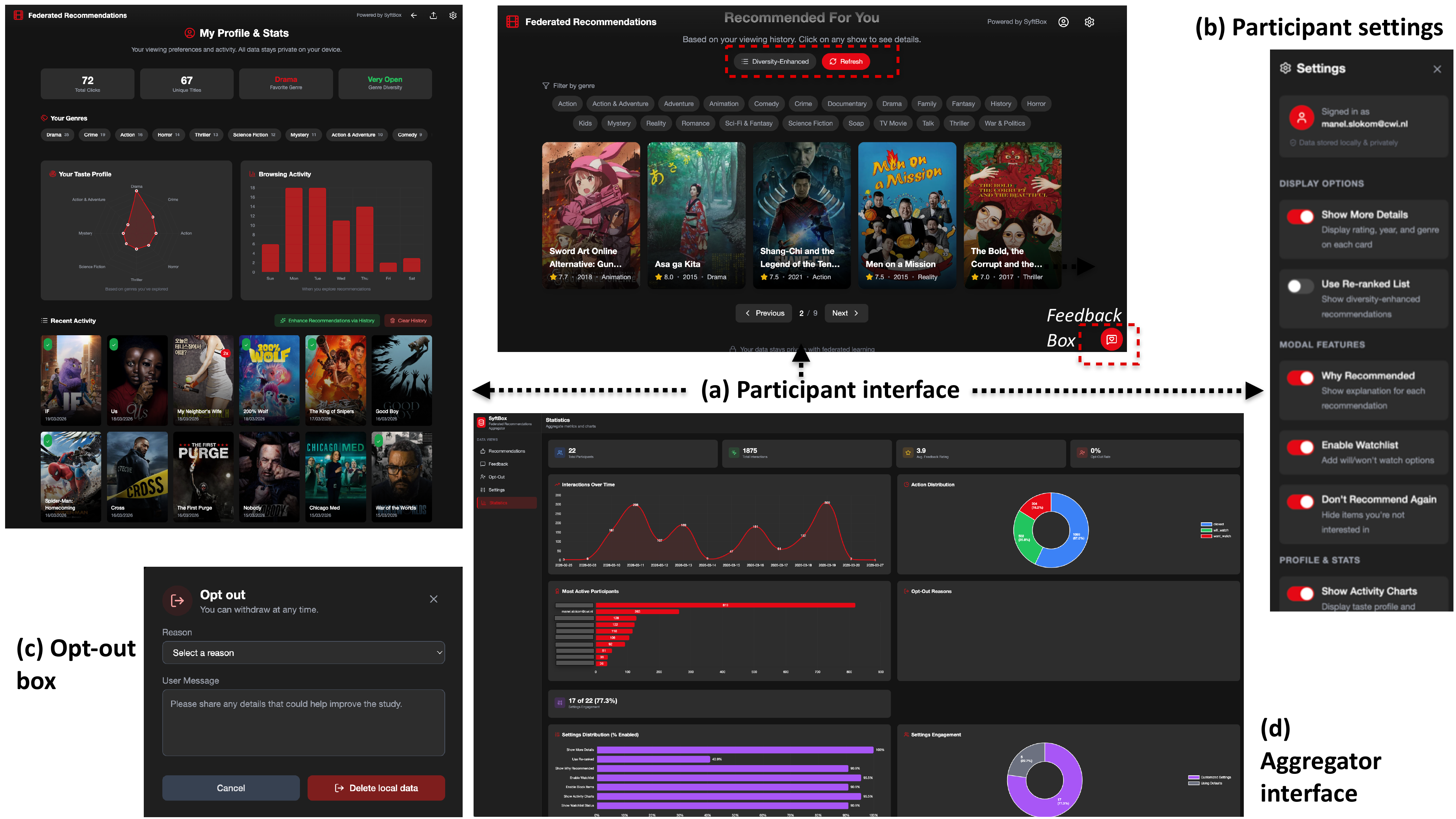}
    \caption{User interface and control mechanisms. \textbf{(a)} Recommendation feed with local filtering and diversity-enhanced mode toggle. \textbf{(b)} Settings panel for granular control over display, explanation, and feedback features. \textbf{(c)} Explicit opt-out mechanism with reasoning collection. \textbf{(d)} Aggregator dashboard showing system-wide engagement and settings adoption patterns.}
    \label{fig:app}
\end{figure*}

We design interfaces for control through action feedback loops rather than passive explanations. 
Our participant and aggregator interfaces are depicted in Figure~\ref{fig:app}.

\paragraph{Ranking objectives.}
Participants toggle between two ranking objectives at any time:
\begin{itemize}
  \item \textbf{Personalization-only:} ranking based on personalization signals.
  \item \textbf{Diversity-enhanced:} 60\% genre-diverse content and 40\% personalized items via Maximal Marginal Relevance (MMR).
\end{itemize}
Mode switches show users how their choice affects rankings.

\paragraph{User agency controls.}
Additional controls include: ``Why recommended,'' ``Don't recommend again,'' watchlist support, activity charts, and local-only mode (disables shared learning). 
These controls enable users to influence optimization objectives and data sharing.

Our guiding principle is \textit{what you see = what you control}.

\paragraph{Demo experience.}
Attendees can (i) interact with the recommendation feed, (ii) toggle ranking mode, (iii) provide negative feedback, and (iv) inspect profile signals and changes. The onsite demo uses mock users without multi-machine federation, while illustrating the aggregation loop.

\section{Implementation and Deployment}
\label{sec:implementation}

We implement local training with BPR~\cite{BPR}, where each participant updates parameters from their interaction history for aggregation. 
A server routine aggregates updates to compute global parameters. 
We apply differential privacy by clipping updates and adding Gaussian noise before aggregation. 
Updated global parameters are then redistributed to participants.

We build both participant and aggregator components as runnable applications, using SyftBox as the aggregation transport layer~\cite{pysyft}.

\paragraph{Deployment setup}
We deployed over 53 days (Jan 25--Mar 21, 2026) with 22 participants across 6 countries and 8807 catalog titles. 
We logged interaction events, settings changes, and optional user feedback.

\paragraph{Ethics.}
The study protocol was reviewed and approved by an institutional ethics committee prior to deployment.
All participants provided informed consent. 
No raw viewing data left user devices; only model updates were shared, protected through secure aggregation and differential privacy mechanisms. 
Interaction data were stored in encrypted form on secure servers.

\paragraph{Metrics.}
We define click-through rate (CTR) as the fraction of displayed recommendations that users click on. 
We collected 1867 recommendations and user satisfaction ratings on a 1--5 scale.
\section{Results: Key findings}

\textbf{Finding 1: Personalization dominates under user choice}
Standard recommendations achieve higher CTR than diversity-enhanced ranking (65.37\% vs.\ 62.07\%, +3.30\%). 
This contrasts with assumptions that diversity universally improves user experience. 
Users prefer personalization by default but value choosing diversity when desired, suggesting diversity works best as a \emph{user-controlled option} rather than a system-imposed constraint.

\begin{figure}[h]
\centering
\includegraphics[width=0.5\textwidth]{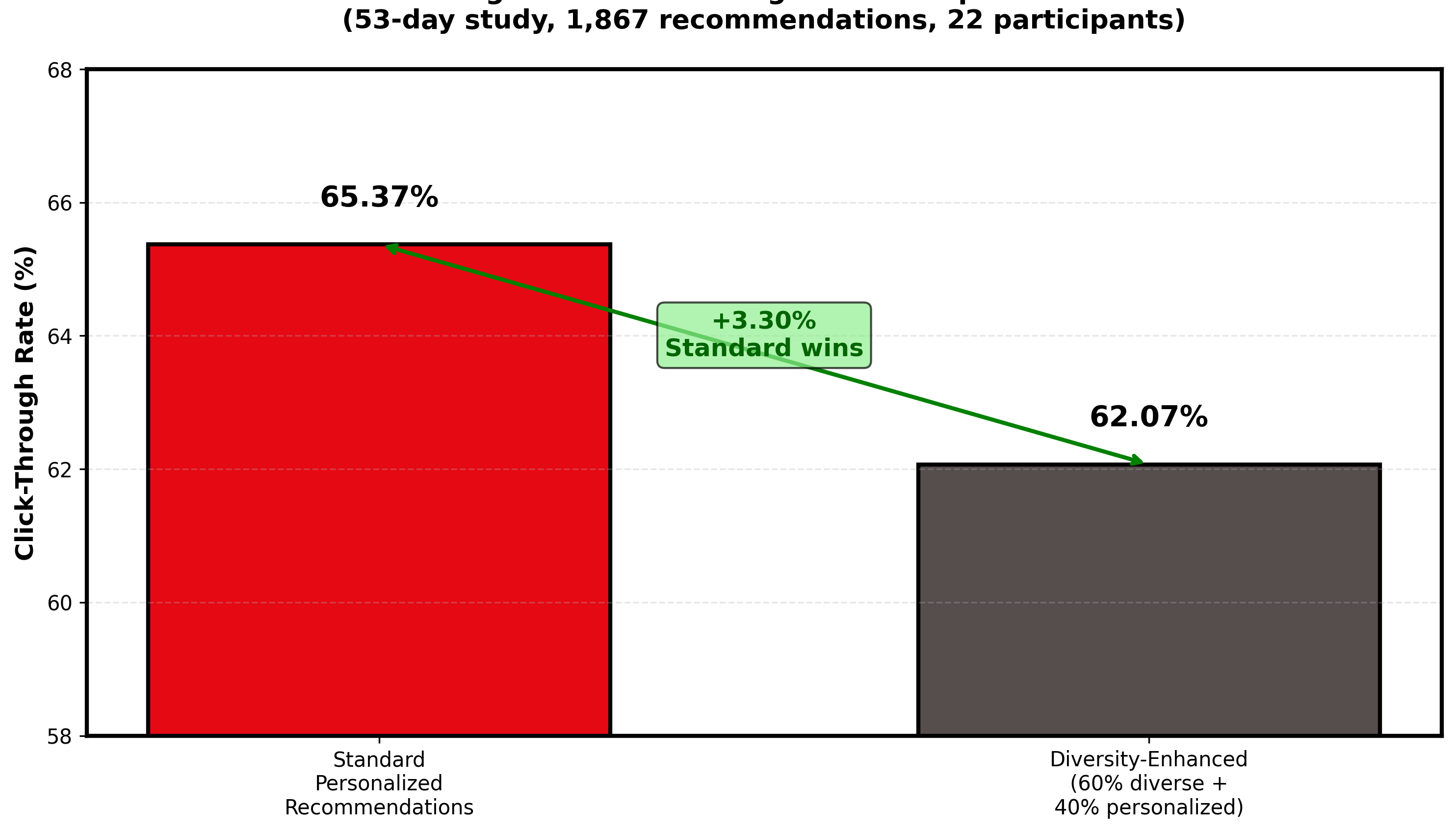}
\caption{CTR comparison: standard personalization outperforms diversity-enhanced ranking by 3.30 percentage points.}
\label{fig:ctr}
\end{figure}

\textbf{Finding 2: Personalization is stable, diversity requires tuning}
Standard recommendations maintain stable CTR (65--67\%) over 53 days. 
Diversity-enhanced ranking starts at 62\% and declines slightly, reflecting real-world non-stationarity and feedback effects. 
Personalization is robust; diversity-oriented objectives require continuous calibration.

\begin{figure}
    \centering
    \includegraphics[width=0.5\textwidth]{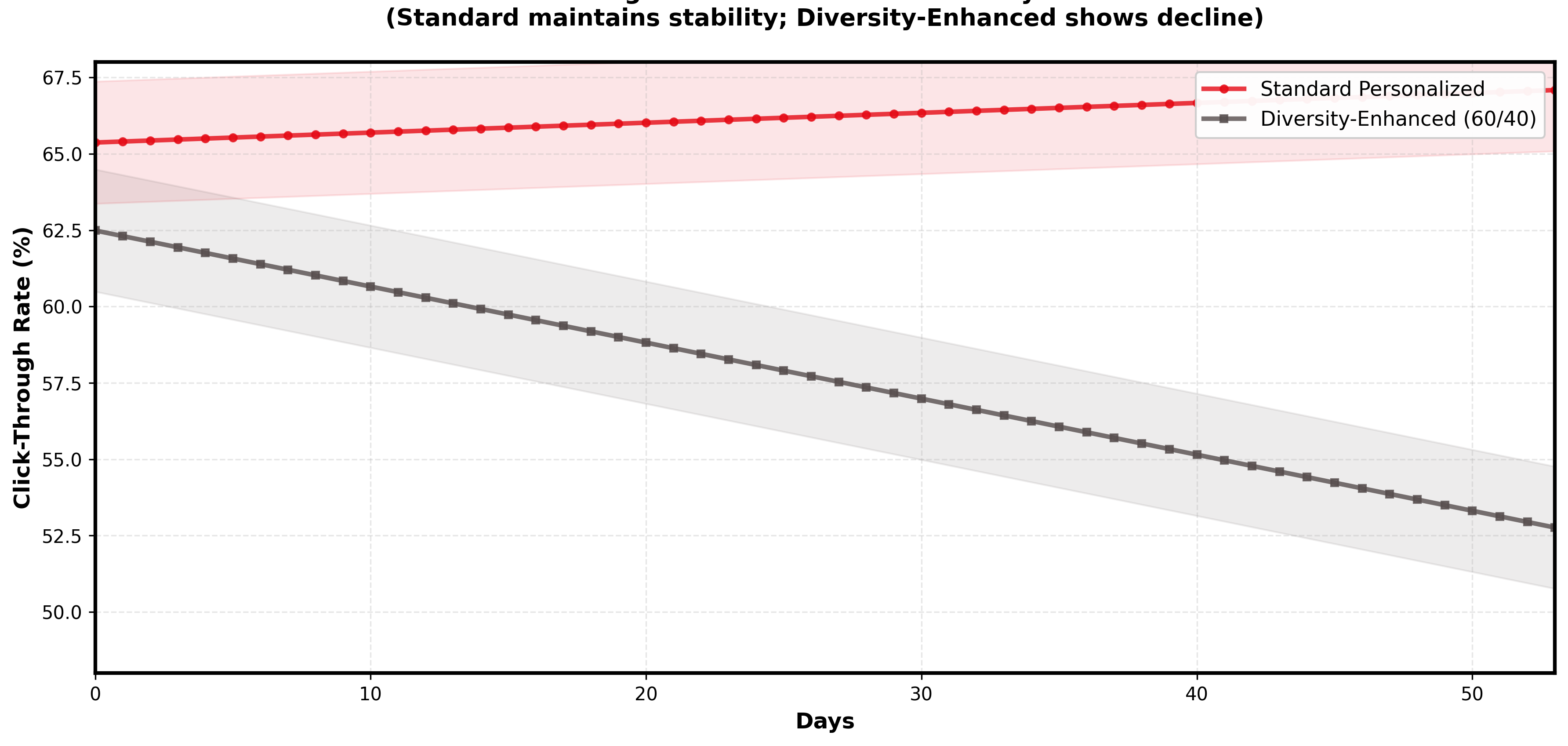}
    \caption{CTR trend over 53 days shows stable personalization versus declining diversity.}
    \label{fig:trend}
\end{figure}

\textbf{Finding 3: Users actively engage with control mechanisms}
Participants made 248 settings changes, submitted 27 feedback entries, and took 79+ ``don't recommend'' actions. 
Average satisfaction was 3.93/5.

Qualitative feedback reveals users develop causal understanding through interaction. 
One participant noted that enabling history-based learning clarified recommendation behavior, showing that interfaces can expose system logic without technical explanation. 
Users reported perceived quality improvements: ``Great, I got some items that I want to watch from the first recommended 50 items, and also from the refreshed items based on what I click'' (User, 5/5), and ``In the second session, there was already a noticeable improvement in the genres suggested'' (User, 4/5).

\textbf{Demo experience.}
Attendees browse recommendations, toggle between ranking modes, provide feedback, and inspect system behavior in real time, experiencing how local control produces immediate, observable effects.




\section{Conclusion}

Our live deployment demonstrates that users favor personalization when given explicit choice, contrasting with assumptions that diversity universally improves user experience. 
Control mechanisms increase trust through visible cause-and-effect relationships, helping users to understand the system's behavior. 
These findings highlight four key insights: (i) live deployment reveals genuine user preferences, (ii) control increases trust, and (iii) diversity works best as a user-selected option rather than enforced constraint.




\section*{Acknowledgment}
The authors would like to thank the 22 participants for their valuable involvement in the experiments. 
Manel's research is supported by the AI, Media \& Democracy Lab. For more information about the lab and its further
activities, visit \url{https://www.aim4dem.nl/}.
\bibliographystyle{ACM-Reference-Format}
\bibliography{paper}

\end{document}